\documentclass[aps,prx,reprint,twocolumn,floatfix,citeautoscript,longbibliography,superscriptaddress]{revtex4}

\usepackage{graphicx}
\usepackage{bm,amsmath,amssymb,mathrsfs,dcolumn}
\usepackage{latexsym}
\usepackage{epsfig}
\usepackage{amsbsy}
\usepackage{array}
\usepackage{setspace}
\usepackage[colorlinks=true, linkcolor=blue, citecolor=blue, urlcolor=blue, linktoc=page, bookmarks=false, pdfstartview={FitH}, pdfborder={0 0 0.0 [3 3]}]{hyperref} 
\usepackage[usenames]{color}
\usepackage[final]{pdfpages}
\usepackage{epstopdf}
\usepackage{subfigure}
\usepackage{units}
\usepackage{url}
\hypersetup{pdfborder=0 0 0,colorlinks=true,citecolor=blue,linkcolor=blue}

\makeatletter

\newcommand{\Rmnum}[1]{\expandafter\@slowromancap\romannumeral #1@}
\makeatother

\begin{document}

\title{Tunneling magnetoresistance in Mn$_2$%
Au-based pure antiferromagnetic tunnel junction}

\author{Xingtao Jia}
\altaffiliation{Corresponding author, E-mail:jiaxingtao@hpu.edu.cn}
\affiliation{School of Physics and Electronic Information
Engineering, Henan Polytechnic University, Jiaozuo 454000, China}
\date{\today }

\author{Hui-Min Tang}
\altaffiliation{Corresponding author, E-mail:hmtang@gxnu.edu.cn}
\affiliation{School of Physical Science and Technology, Guangxi
Normal University, Guilin 541001, China.}

\author{Shi-Zhuo Wang}
\affiliation{School of Physics and Electronic Engineering, Zhengzhou
University of Light Industry, Zhengzhou 450002, China}

\begin{abstract}
Antiferromagnetic (AF) spintronics is merit on ultra-high operator speed and stability in the presence of magnetic field. To fully use the merit, the device should be pure rather than hybrid with ferromagnet or ferrimagnet. For the magnetism in the antiferromagnet is canceled by that of different sublattices, breaking the symmetry in the material can revive the native magnetism, which can be detected by the magnetoresistance (MR) effect. Achieving noticeable MR effect in the pure AF device is difficult but essential for the AF spintronic applications.
Here, we study the tunnel magnetoresistance (TMR) effect in the Nb$/$Mn$_2$Au$/$CdO$/$Mn$_2$Au$/$Nb pure AF magnetic tunnel junctions (AF-MTJs) based on a first-principle scattering theory. Giant TMRs with order of $1000\%$ are predicted in some symmetric junctions, which is originated from the interfacial resonance tunneling effect related with the $k_{||}$ dependent complex band structures of CdO and Mn$_2$Au in companion with the enhanced spin polarization of the interfacial magnetic atoms. The effect of voltage bias and interfacial disorder such as Oxygen vacancy, Manganese vacancy, and Manganese-Cadmium exchanges at Mn$_2$Au$/$CdO interfaces are studied also.
Our studies suggest Nb$/$Mn$_2$Au$/$CdO$/$Mn$_2$Au$/$Nb AF-MTJs promising material for AF spintronic application, and rocksalt CdO a potential symmetry filtering material for spintronic applications.
\end{abstract}

\maketitle
\section{Introduction}

As one candidate of the next generation electronics, antiferromagnetic (AF) spintronics is merit on ultra-high speed spin dynamics up to THz and insensitivity to magnetic field,\cite{jungwirth2016antiferromagnetic,baltz2018antiferromagnetic} which is promising as the magnetic memories and magnetic sensors, and also demonstrates potential application for brain-inspired computation. To fully use the merit, the AF spintronic devices should avoid to use ferromagnet (FM) and ferrimagnet as reference layer or pinning layer. That is, which should be pure rather than hybrid. Magnetoresistance (MR) effect is general in the magnet. The no-relativistic MR effect is always larger than the relativistic one. The MR effect in the pure AF spin valves (AF-SVs) is different in principle from that in the ferromagnetic one,\cite{nunez2006theory} which is small in general and hard to meet the commercial demand. By taking advantage of the perfect spin filter effect of the MgO barrier, tunneling MR (TMR) effect over several hundreds percent can be found in some hybrid AF tunnel junctions(AF-MTJs) with FM as spin polarizer and antiferromagnet (AFM) as active layer. \cite{jia2020giant,jia2017structure}
Recent studies predict giant TMR up to several thousand percent in the sandwich-type van der Waals (vdW) magnetic tunnel junctions,\cite{PhysRevB.103.134437,yan2020significant,PhysRevB.104.144423,PhysRevApplied.16.024011,PhysRevApplied.17.034030} which is related with the magnetic state dependent band structure. Unconventional colossal angular MRs are found in the semiconductor-type AFM EuTe$_2$ with space-time inversion symmetry-broken\cite{PhysRevB.104.214419}, where the metal-insulator transition (MIT) induced by the magnetic field should be responsible for. Large TMR as high as $300\%$ are predicted first-principlely in the Mn$_3$Sn based AF-MTJs originated from the spin-spliting in the moment space\cite{PhysRevLett.128.197201}, and TMR about $100\%$ in RuO$_2$ based AF-MTJ originated from spin-momentum coupling\cite{PhysRevX.12.011028}.

Electric current is preferred to manipulate the AFM,\cite{0957-4484-29-11-112001,gomonay2014spintronics,jungwirth2016antiferromagnetic,baltz2018antiferromagnetic}
which is compatible to the state-of-art semiconductor technology. When the current flows through the AF metal with broken inversion symmetry, the N{\'e}el spin-orbit torque (NSOT)\cite{Wadley587,bodnar2018writing,PhysRevLett.120.237201,PhysRevApplied.9.054028} and spin orbital torque (SOT)\cite{deng2022} can manipulate the N{\'e}el order with a current density around $10^{6}~\unit{A}/\unit{cm}^2$,
which is comparable to the current induced spin transfer torque (STT) in
the well-studied MgO-based ferromangetic tunnel junctions (F-MTJs).\cite{butler2001,mathon2001theory,ikeda2010perpendicular}
Current flowing though the noble metal can induce spin Hall effect (SHE), which can be used to manipulate the N{\'e}el vector of AFM.\cite{jia2020spin} Antidamping torque from the SHE effect can be used to control the N{\'e}el order efficiently by control the epitaxial orientation.\cite{zhou2019fieldlike} Generally, the switching current in AFM using SHE is orders in magnitude larger than that using NSOT effect for the lower efficiency of the former, which can be considerable enhanced by topological surface states.\cite{hai2020spin}

The spin dynamics in the AFM can be depicted by a coupled Landau-Lifshitz-Gilbert (LLG) equation.\cite{jungwirth2016antiferromagnetic,baltz2018antiferromagnetic}
Therein, the time scale of the spin dynamics is dependent on the exchange field between the magnetic moments. The spin dynamics in the synthetic 2D AFM is more similar to that in the spin valve with two FMs sandwiched by non-magnetic materials. The synthetic 2D AFM can show giant MR effect,\cite{PhysRevB.103.134437,yan2020significant,PhysRevB.104.144423,PhysRevApplied.16.024011,PhysRevApplied.17.034030} but cannot be considered as AF spintronic device. To fully realize the merits of the AFM, all the magnetic materials in the AF spintronic device should be intrinsic AFM. 
Unfortunately, reports about the MR effect in the pure AF spintronic device is rare. 
 Seeking a pure AF device with large MR effects and good accommodation with the current-in-art semiconductor architecture is necessary and ungent for the AF spintronic applications.

Sizeable MR effect in the hybrid AF spintronic device is believed to be related with the localized interfacial states,\cite{jia2017structure,jia2020spin} which demand that the interfaces should be as clean as possible. The interfacial disorders, including the atomic disorder and spin disorder, can destroy the interfacial states, which is key factor for sizeable MR effect in the hybrid AF-MTJs.\cite{jia2017structure,jia2020giant} To realize ideal interfaces, epitaxial contact between the AF metal and non-magnetic insulator should be favorable. Here, we focus on epitaxial Mn$_2$Au$/$CdO$/$Mn$_2$Au pure AF-MTJs with ideal interface. Tetragonal Mn$_2$Au shows high N{\'e}el temperature above $1000\unit{K}$.\cite{barthem2013revealing} Current induced NSOT effect is efficient to manipulate the N{\'e}el vector of metallic Mn$_2$Au for the broken inversion symmetry structure.\cite{Wadley587} Cubic CdO shows rocksalt (NaCl) structure with lattice parameter $a=4.69~\unit{\AA}$, which can match well with the tetragonal Mn$_2$Au by rotation of $45$ degree. For the current induced STT is inefficiency to manipulate the N{\'e}el order of the AFM in the perpendicular magnetic structure, three terminals structure using NSOT or SHE effect is more favorable. In these case, the Mn$_2$Au layer thickness should be as thinner as possible. Bcc Nb or Ta with lattice parameter $a=3.3~\unit{\AA}$ can be used as leads, which can match well with Mn$_2$Au also, and the former is more favorable for low price.

\begin{figure}[tbp]
\centering
\includegraphics[width=8.6cm]{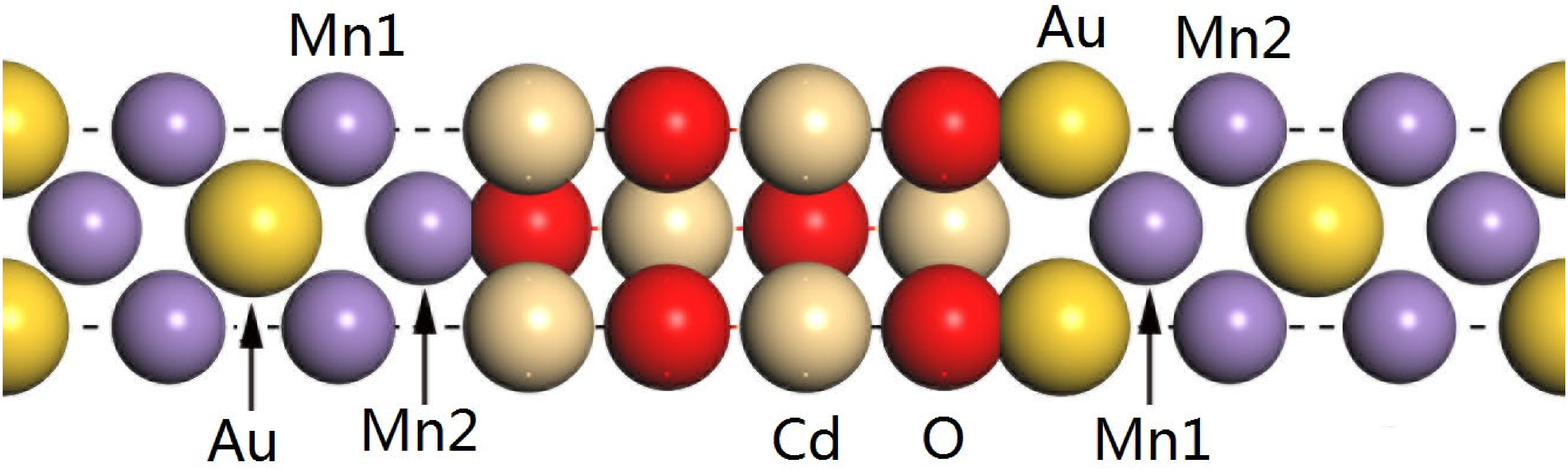}
\includegraphics[width=8.6cm]{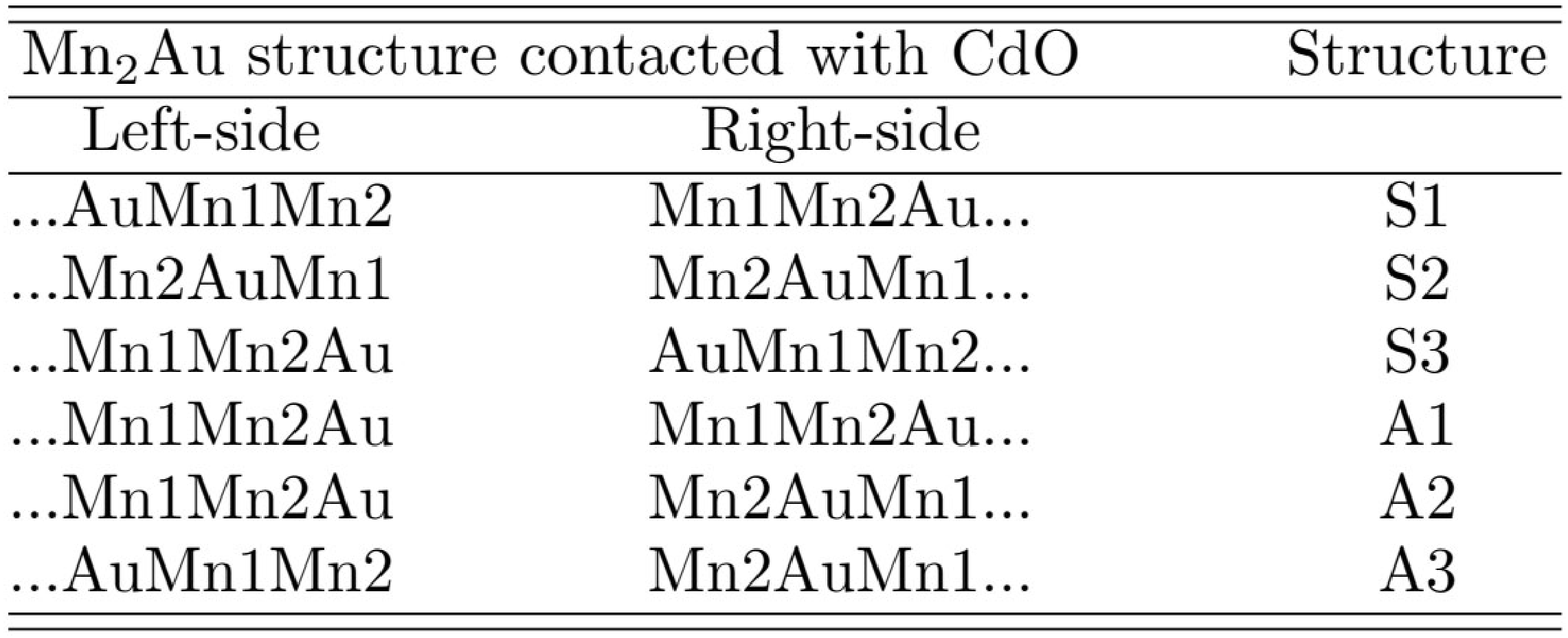}
\caption{Schematic Mn$_2$Au$/$CdO$/$Mn$_2$Au(001) multilayer used in the study. Therein, Mn$_2$Au follows a ...ABCABC... structure along the [001] direction with A, B and C standing for Au, Mn at site one (Mn1), and Mn at site two (Mn2), respectively. According to the termination atoms of Mn$_2$Au contacted with CdO, we define three symmetric structures named as S1, S2 and S3 and three asymmetric structures named as A1, A2 and A3.}
\label{fig1}
\end{figure}

\section{Methods}

A two-terminal structure with Mn$_2$Au$/$CdO$/$Mn$_2$Au sandwiched between two semi-infinite Nb leads was used to study the MR effect in the study. The stacking of Mn$_2$Au$/$CdO$/$Mn$_2$Au multilayer is shown in Fig. \ref{fig1}. We firstly carry Density functional theory (DFT) calculations based on the plane wave function to found the stable structure, and then carry tight-binding linear muffin-tin orbital (TB-LMTO) calculations to get the potentials of the system, and finally transfer the potentials into a Wave-Function-Matching (WFM) package to calculate the spin transports.
\subsection{Structure calculations}
We use DFT calculations based on the projector augmented wave (PAW) potential and PBE exchange-correlation function\cite{vasp} to find stable structures. A slab structure containing 6 Ls CdO and 6 Ls Mn$_2$Au is used to study the Mn$_2$Au$/$CdO(001) interface. During calculations, we expand CdO to match with Mn$_2$Au, and all parameters are fixed to their bulk state and the only free parameters is the distance
between Mn$_2$Au and CdO. A cutoff energy of $600~\unit{eV}$ is taken for all calculations. DFT calculations indicate the O termination (of CdO) cases are energy stable. The total energies are -129.092, -130.038, and -130.047 $\unit{eV}$ for the Au, Mn1, and Mn2 atoms (of Mn$_2$Au) bonding with O atoms (of CdO) with bonding lengths around 2.41, 2.03 and $1.93~\unit{\AA}$, and we denote the structures as "Au termination", "Mn1 termination" anbd "Mn2 termination" in the study, respectively. That is, the Mn$_2$Au$/$CdO interface with Mn2 termination structure most energy favorable. For the Nb$/$Mn$_2$Au interface, the DFT calculations indicate that the structure with Mn2 atoms (of the Mn$_2$Au layer) sited above the second monolayer Nb atoms (of the Nb layer) is more energy favorable.
\subsection{potential calculations}
The potentials used in the transport package are obtained via self-consistent calculations performed using TB-LMTO surface Green's function method with a coherent potential approximation (CPA) to deal with imperfects.\cite{turek1997electronic} For Nb, the atomic sphere has radius 1.639 $\unit{\AA}$ whose space fills the bcc lattice. For Mn$_2$Au, same atomic radius of 1.555 $\unit{\AA}$ are used for both Au and Mn atoms to fill the tetragonal lattice. Inside CdO, we take radius 1.492 $~\unit{\AA}$ for oxygen atoms and 1.237 $\unit{\AA}$ for Cd atoms, and vacuum sphere of radius 0.796 $\unit{\AA}$ is added at the center of the cube with 4 O atoms and 4 Cd atoms to fill the space. The direct band gap of CdO present at $\Gamma$ point is around 2.85 $\unit{eV}$ using modified Becke-Johnson (mBJ)\cite{tran2009accurate} potential within local spin density approximation (LDA), and the conduction band minimum (CBM) is around $1.3~\unit{eV}$ above the Fermi energy. Indirect band gap is around $1.5~\unit{eV}$ with the valance band maximum (VBM) localized at L point (see Fig.\ref{fig3}B). The band structure is consistent with the experimentals\cite{liu2003synthesis,chandiramouli2013review}. Three empty spheres are introduced to fill the Mn$_2$Au$/$CdO interface during the self-consistent and spin transport calculations. Therein, two vacuum spheres with radius r$_1$ are inserted exactly above the vacuum spheres inside the CdO, and a vacuum sphere with radius r$_2$ is added exactly above the Cd atom. The former are considered to belong to the CdO layer, and the positions of the latter is arranged to minimize overlap. For the Mn2 termination structure, both r$_1$ and r$_2$ are set to $0.653~\unit{\AA}$. For the Mn1 termination structure, both r$_1$ and r$_2$ are set to $0.796~\unit{\AA}$ and $0.668~\unit{\AA}$, respectively. For the Au termination structure, both r$_1$ and r$_2$ are set to $0.796~\unit{\AA}$ and $0.956~\unit{\AA}$, respectively.
\subsection{transport calculations}
The spin transport calculations were carried out via a first-principle WFM method\cite{wang2008first} for the ideal junctions. For cases with interfacial disorders, the vertex correction\cite{ke08} is used to average the configurations. The N{\'e}el order vectors of Mn$_2$Au are defined by the direction of the magnetic moments of the first Mn atom contacted with CdO. The left-side N{\'e}el order was fixed as reference, and the right-side N{\'e}el order set free. A 1600$\times $1600 \textit{k}-mesh for most structures was used to sample the 2D Brillouin zone (BZ) to ensure good numerical convergence for the ideal epitaxial structures.

\section{Results and Discussion}

The first-principle density-functional theory (DFT) calculations indicate that the Mn$_2$Au$/$CdO interface with O atoms (of CdO) bonding with Mn2 atoms (of Mn$_2$Au) is most energy stable. We denote the structure as "Mn2 termination", and pay more attention to the structure in the study. Comparatively, metastable interfaces with O atoms bonding with Au and Mn1 atoms are considered less.  
According to the types of Mn$_2$Au$/$CdO interface, there are six structures in the Mn$_2$Au$/$CdO$/$Mn$_2$Au multilayer, as shown in Fig. \ref{fig1}. Therein, the S1 structure is more energy favorable, and we focus on the structure in the study.
The tetragonal Mn$_2$Au shows fourth-order in-plane anisotropy with hard axis along $z$ direction and easy axis along $\theta=90^{\circ}$, $\phi=\pm45,$ and $\pm135^{\circ}$.\cite{shick}
According to the relative angle $\Theta$ between the N{\'e}el vector of the reference (left) side ($n_L$) and the free (right) side ($n_R$), there are three stable magnetic states labeled as the parallel (P), perpendicular (PP), and antiparallel (AP) states, corresponding to $\Theta=0, \pi/2$, and $\pi$, respectively. We define the angular-dependant TMR as TMR$(\Theta)=R(\Theta)/R(0)-1$ with resistance $R=1/G$ with conductance $G=(\unit{e}^{2}/\unit{h})Tr({tt^{\dag}}) $, where $t$ is the transmission part of the scatter matrix $S$.

\begin{figure*}[tbp]
\centering
\includegraphics[width=8.6cm]{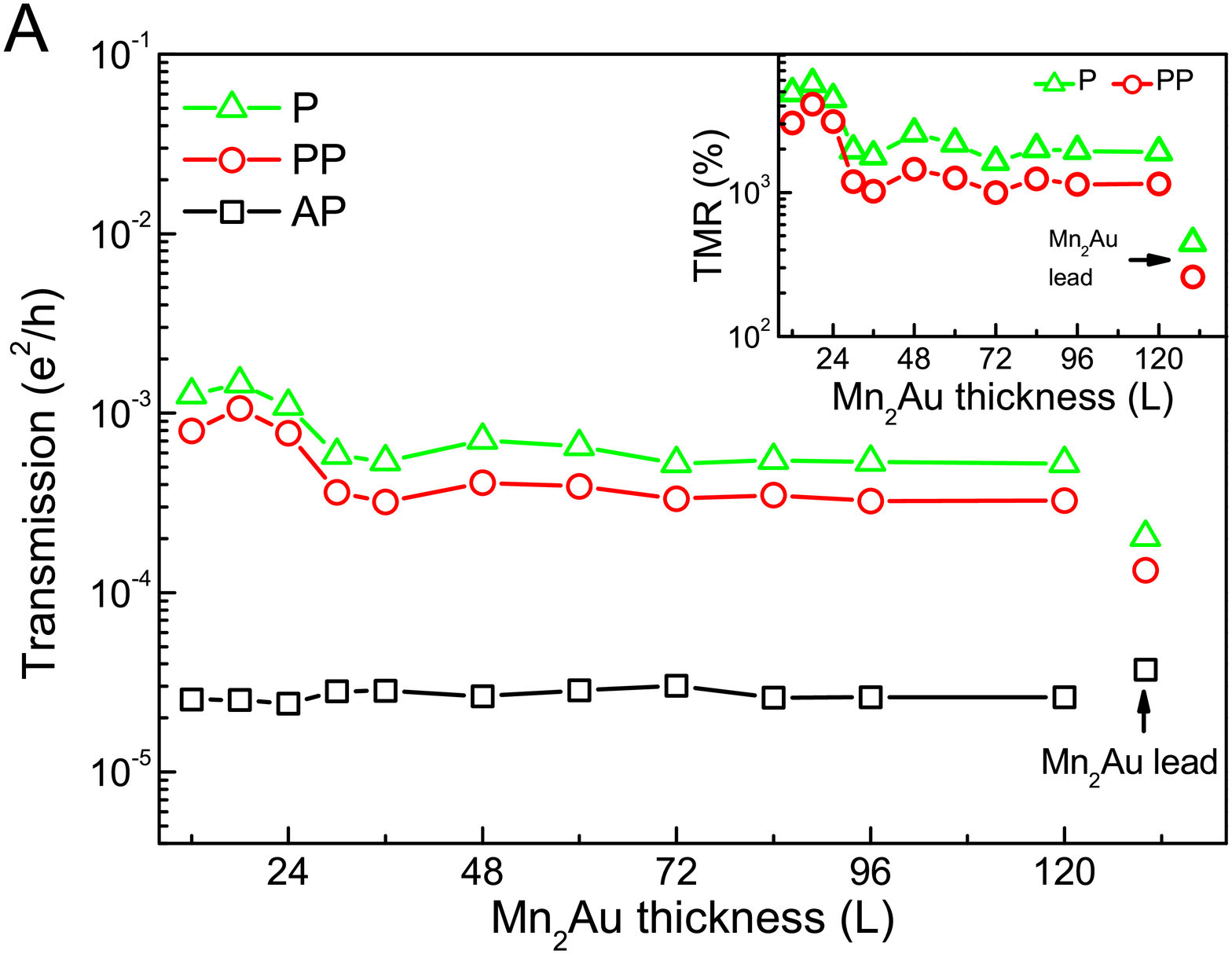}
\includegraphics[width=8.6cm]{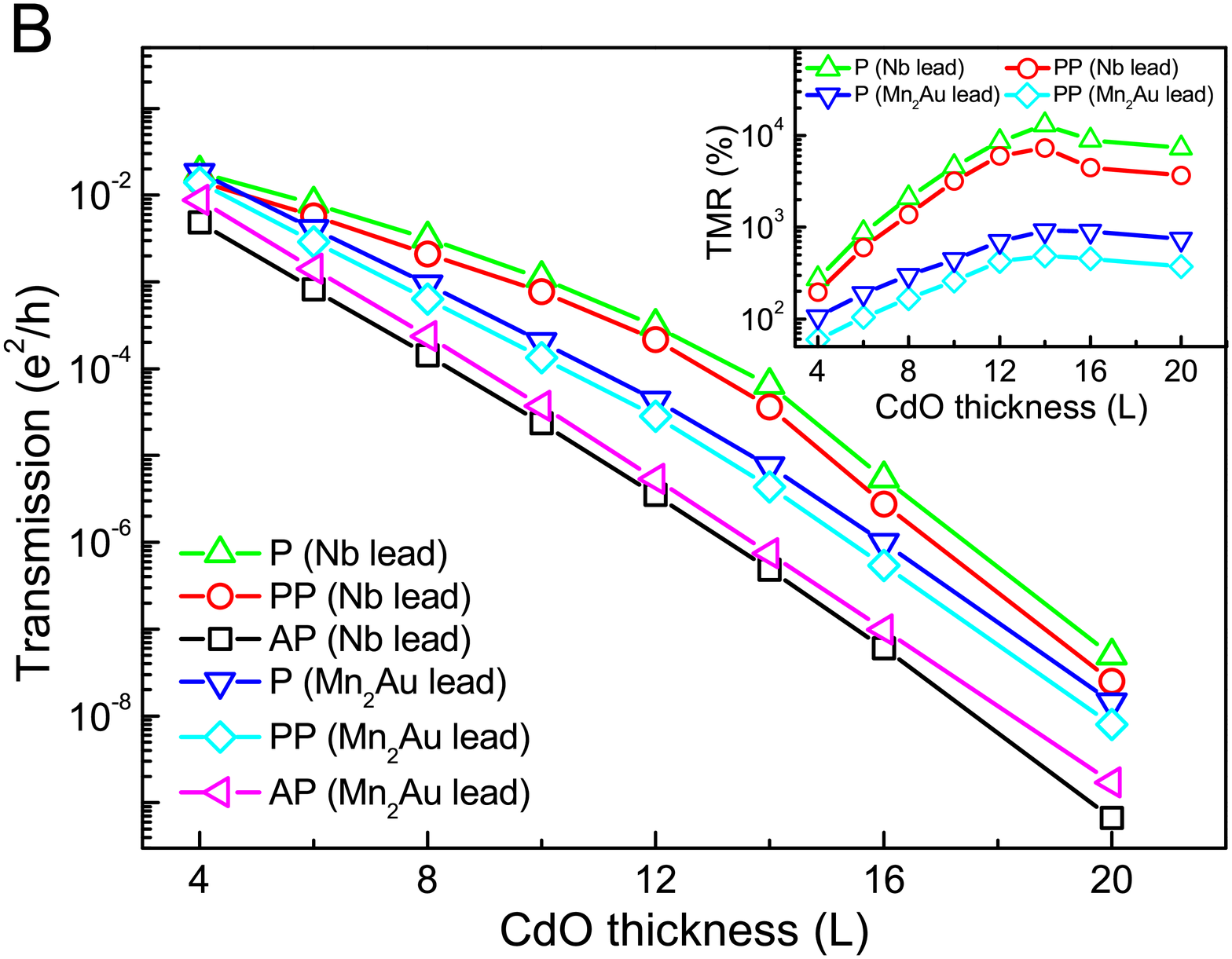}
\includegraphics[width=17cm,bb=0 0 3000 1300]{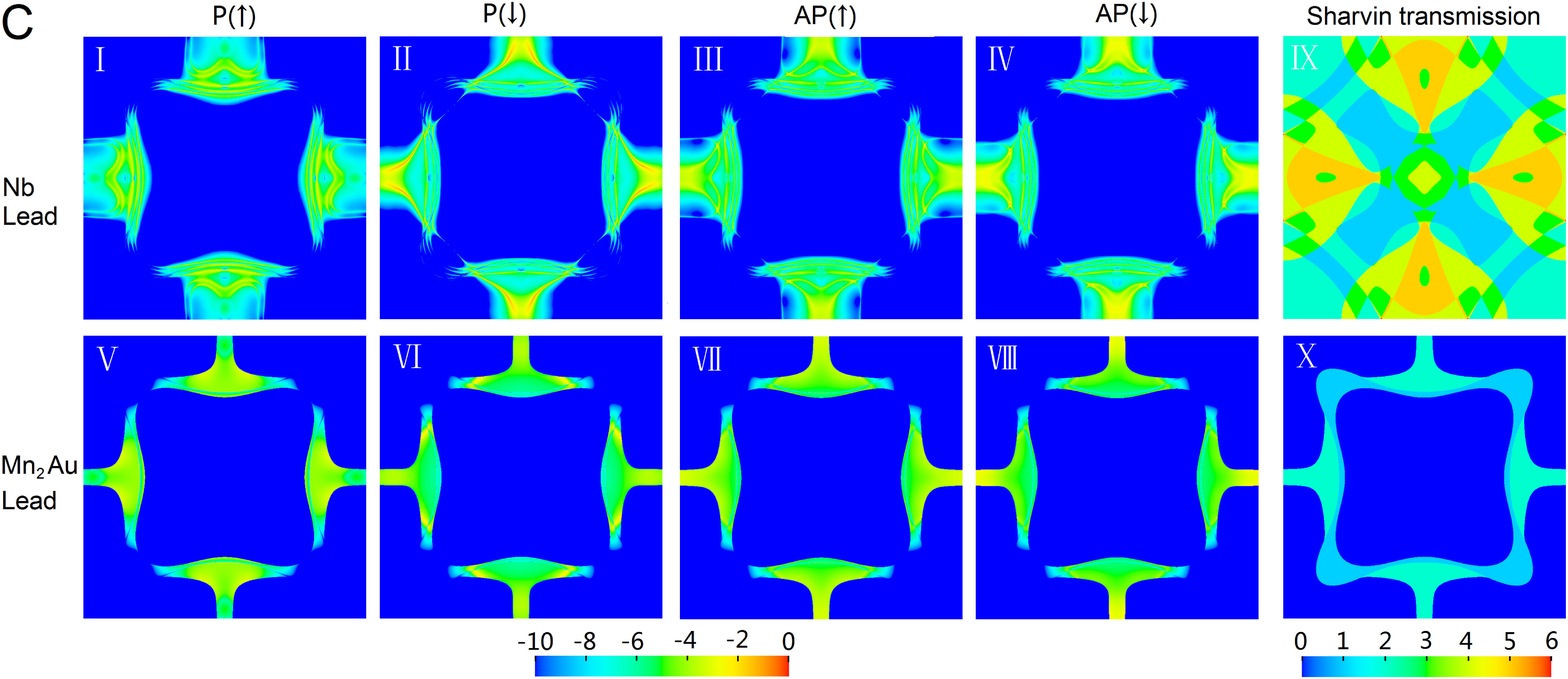}
\caption{(A) Mn$_2$Au thickness dependency and (B)
barrier thickness dependency of the spin transmissions and TMRs of the ideal Nb$/$Mn$_2$Au$/$CdO$/$Mn$_2$Au$/$Nb and
Mn$_2$Au$/$CdO$/$Mn$_2$Au pure AF-MTJs with S1 structure. The CdO barrier
thickness is set 10 Ls in (a), and both the reference
and free Mn$_2$Au layers are set 24 Ls in (b) for the "Nb lead" junctions. (C) $k_{||}$ resolved transmission coefficient in the 2D BZ of the
ideal (\Rmnum{1}-\Rmnum{4}) Nb$/$Mn$_2$Au(24)$/$CdO(10)$/$Mn$_2$Au(24)$/$Nb(001) and
(\Rmnum{5}-\Rmnum{8}) Mn$_2$Au$/$CdO(10)$/$Mn$_2$Au(001) pure AF-MTJs with S1 structure
at Fermi energy, and Sharvin transmission of (\Rmnum{9}) bcc Nb and (\Rmnum{10})
tetragonal Mn$_2$Au along the [001] direction.}
\label{fig2}
\end{figure*}

\subsection{TMRs in the AF-MTJs}

Firstly, we take a look at the thickness dependency of the TMRs. Figure \ref{fig2}A and B give the spin transmissions and TMRs in the ideal Nb$/$Mn$_2$Au(12-120)$/$CdO(4-20)$/$Mn$_2$Au(12-120)$/$Nb pure AF-MTJs with S1 structure, where the numbers in the bracket indicate the thickness in atomic layers (Ls). We name these junctions using Nb leads as "Nb lead" junctions. Firstly, we vary the thickness of Mn$_2$Au while fix 10 Ls CdO as shown in Fig. \ref{fig2}A. Therein, the spin transmissions of the AP structure are less sensitive to the thickness of the Mn$_2$Au region, but about one order in magnitude smaller than that of the very very thicker Mn$_2$Au junction, which can be considered as using Mn$_2$Au as leads and we name it as "Mn$_2$Au lead" junction. The spin transmissions of the AF-MTJs with P and PP magnetic states increase firstly and then decrease with peak values present around 18 Ls Mn$_2$Au, with the peak values around threefold larger than that of the junction with 120 Ls Mn$_2$Au, and around one order in magnitude larger than that of the "Mn$_2$Au lead" junction. Considering the metallicity of tetragonal Mn$_2$Au,\cite{jia2020giant} the huge sensitivity of the spin transmission in the AF-MTJs with respect to the Mn$_2$Au thickness is unusual. The TMR studies indicate giant TMRs with range from $1600\%$ to $5600\%$ in the AF-MTJs as Mn$_2$Au vary from 12 to 120 Ls, which is about one order in magnitude larger than that of the "Mn$_2$Au lead" junction, as shown in the inset of Fig. \ref{fig2}A. The TMRs in these junctions are comparable to the specular TMR effect found in the well-studied MgO-based F-MTJs.\cite{butler2001,mathon2001theory,ke10,jiaprl11} Moreover, the spin transmission of the AF-MTJs follows the simple trigonometric functions, and the TMR effect of the PP magnetic states are around $50\%$ of that of the P states. 

The spin transmissions in both the "Nb lead" and "Mn$_2$Au lead" AF-MJTs decrease exponentially as the CdO thickness increases, with considerable deviation in that with P and PP magnetic states when the barriers are thinner, as shown in Fig. \ref{fig2}B. Here, we fix 24 Ls Mn$_2$Au for the "Nb lead" junctions during the calculations. As the CdO thickness increases, the TMRs of both types junctions increase quickly initially, and then slow down, and finally decrease slowly with peak around 14 Ls CdO. For the junctions with P magnetic state, the lest and largest TMRs are $270\%$ (in junction with 4 Ls CdO) and $13000\%$ for the "Nb lead" junctions, which are $105\%$ and $920\%$ for the "Mn$_2$Au lead" junctions. When the CdO is thicker, the TMRs of the "Nb lead" junctions is about one order in magnitude larger than that of the "Mn$_2$Au lead" junctions. The giant difference indicates the presence of spin filtering effect in the "Nb lead" junctions.

Following to the two-current model, the current flows through the barrier should be spin polarized to induce MR effect. In the limit of non-relativistic case, the current flowing through the AF metal Mn$_2$Au is hard to spin-polarize. However, the spin transmissions and TMRs calculations above indicate that the Mn$_2$Au layers are almost half-metallic, which maybe relate with the interface effect as found in the Fe$/$MgO$/$Ag$/$Mn$_2$Au\cite{jia2020giant} hybrid AF-MTJs.

\begin{figure*}[tbp]
\centering
\includegraphics[width=8.0cm]{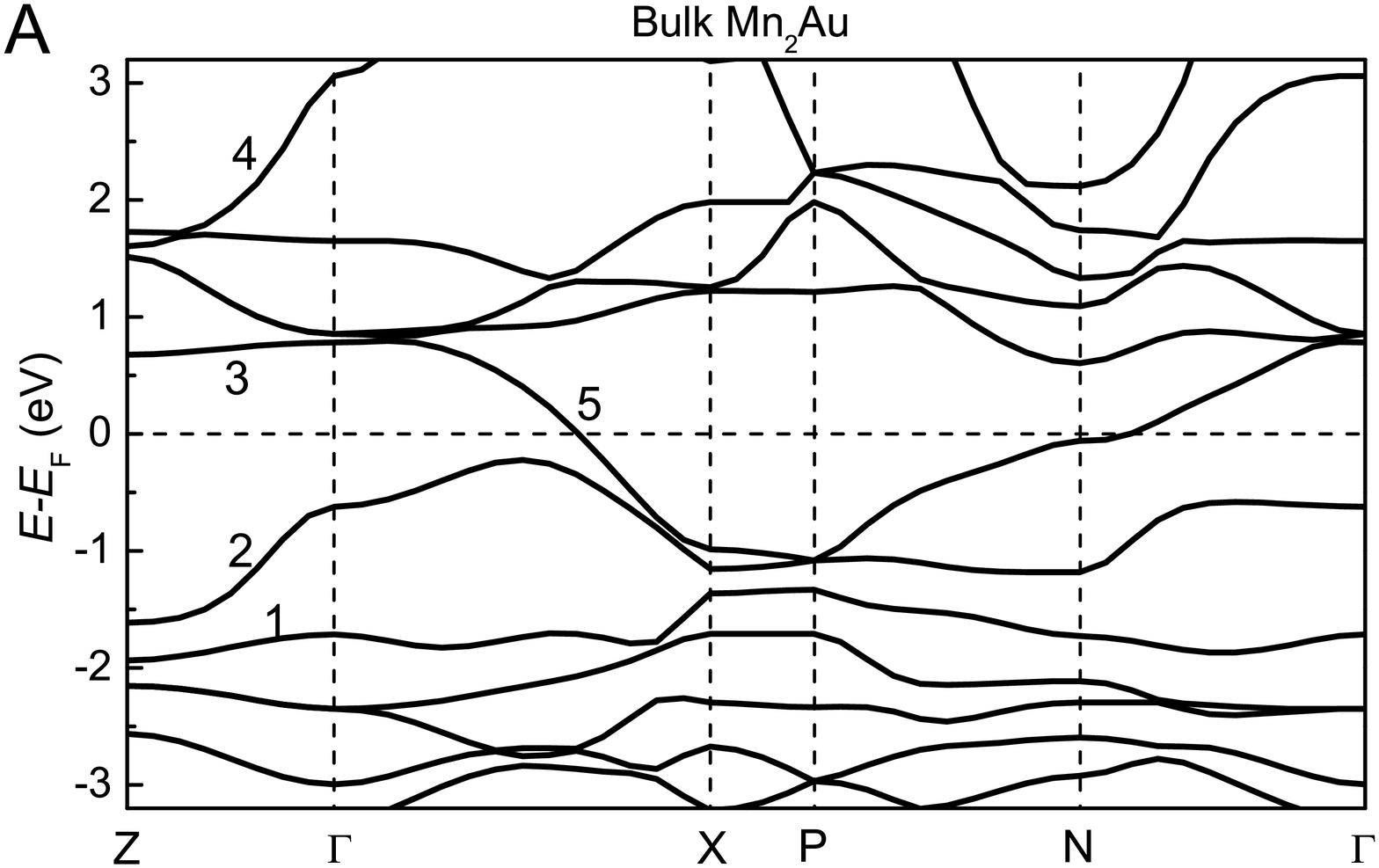}
\includegraphics[width=8.0cm]{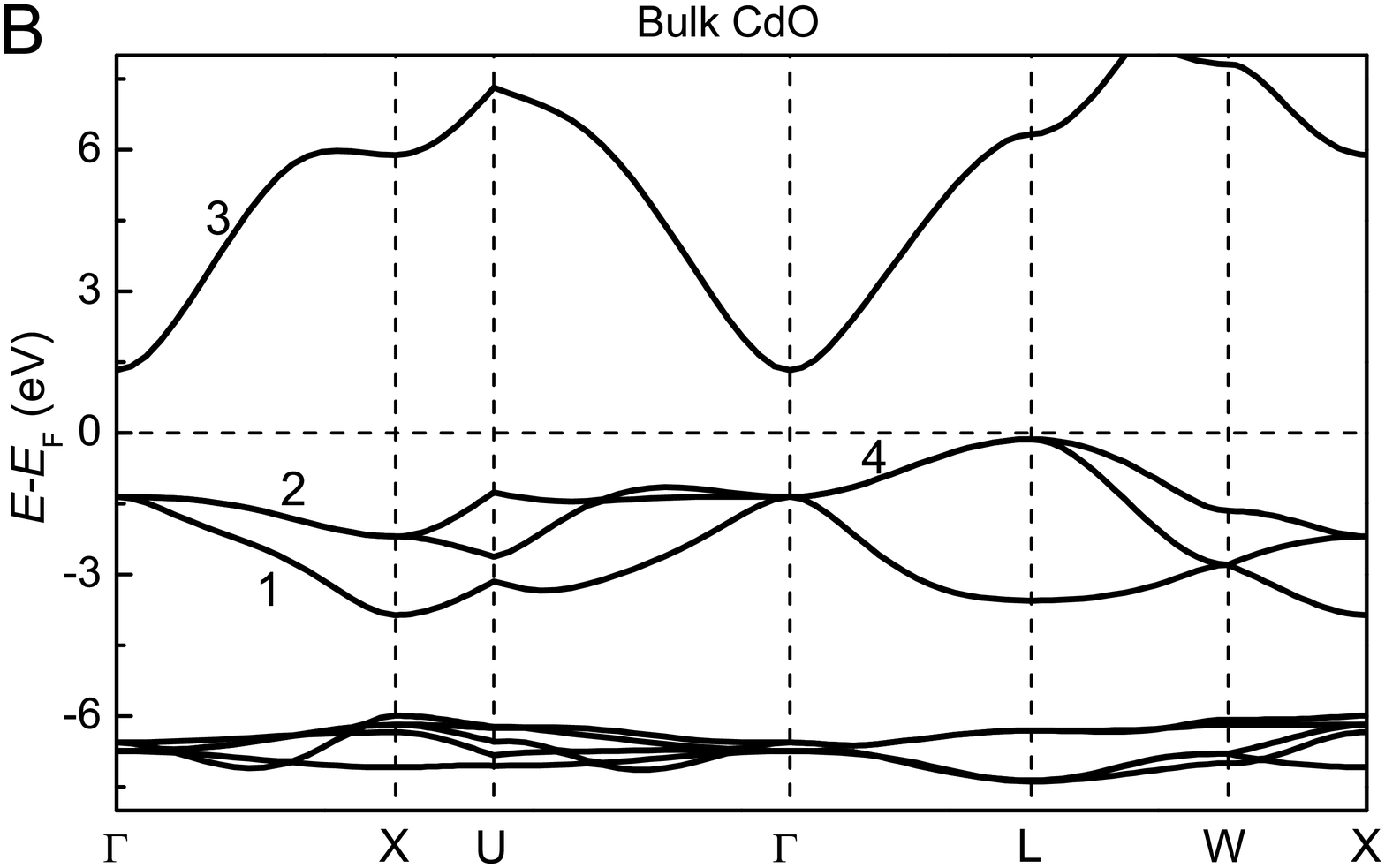}
\includegraphics[width=8.0cm]{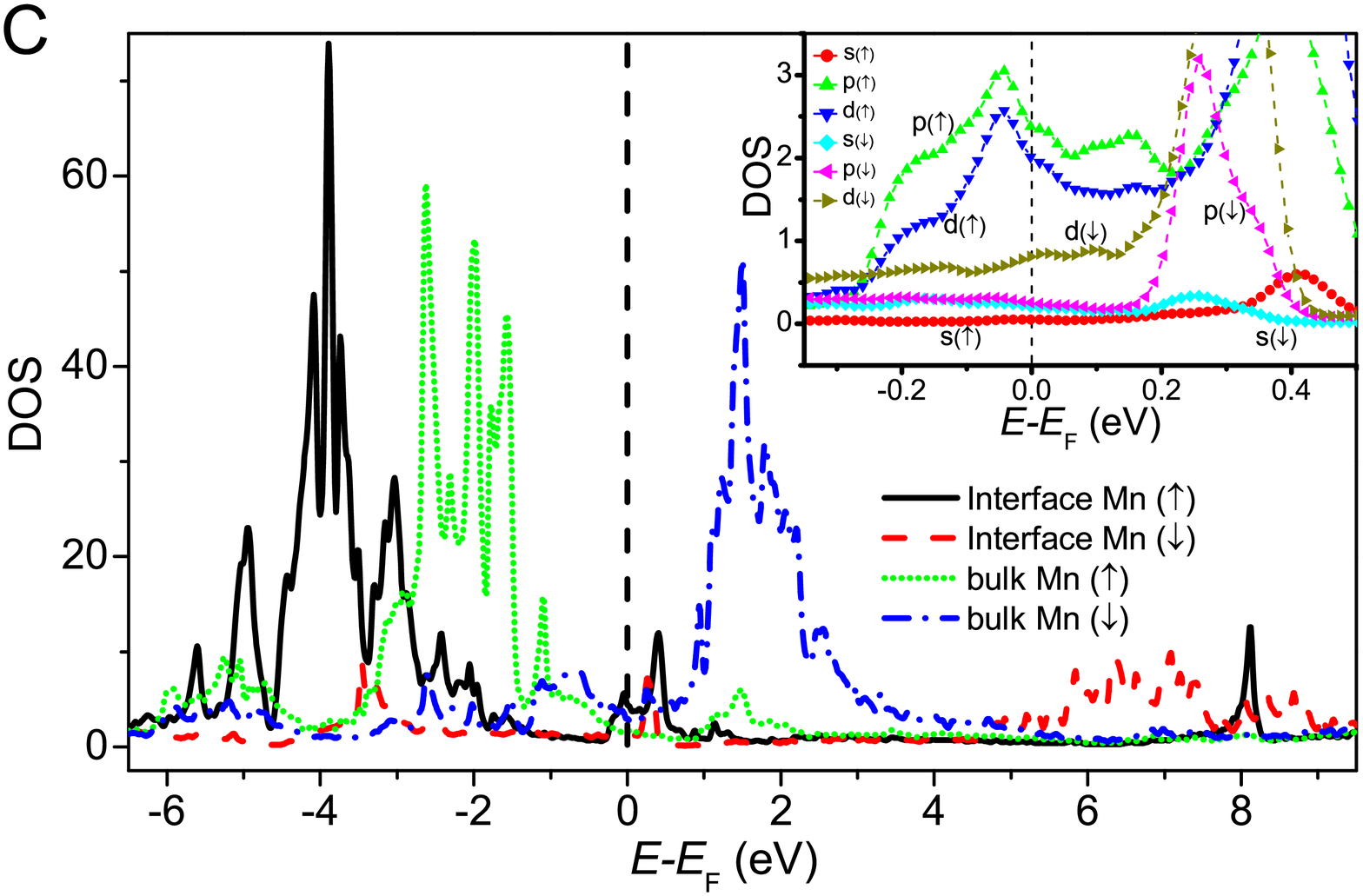}
\includegraphics[width=8.0cm]{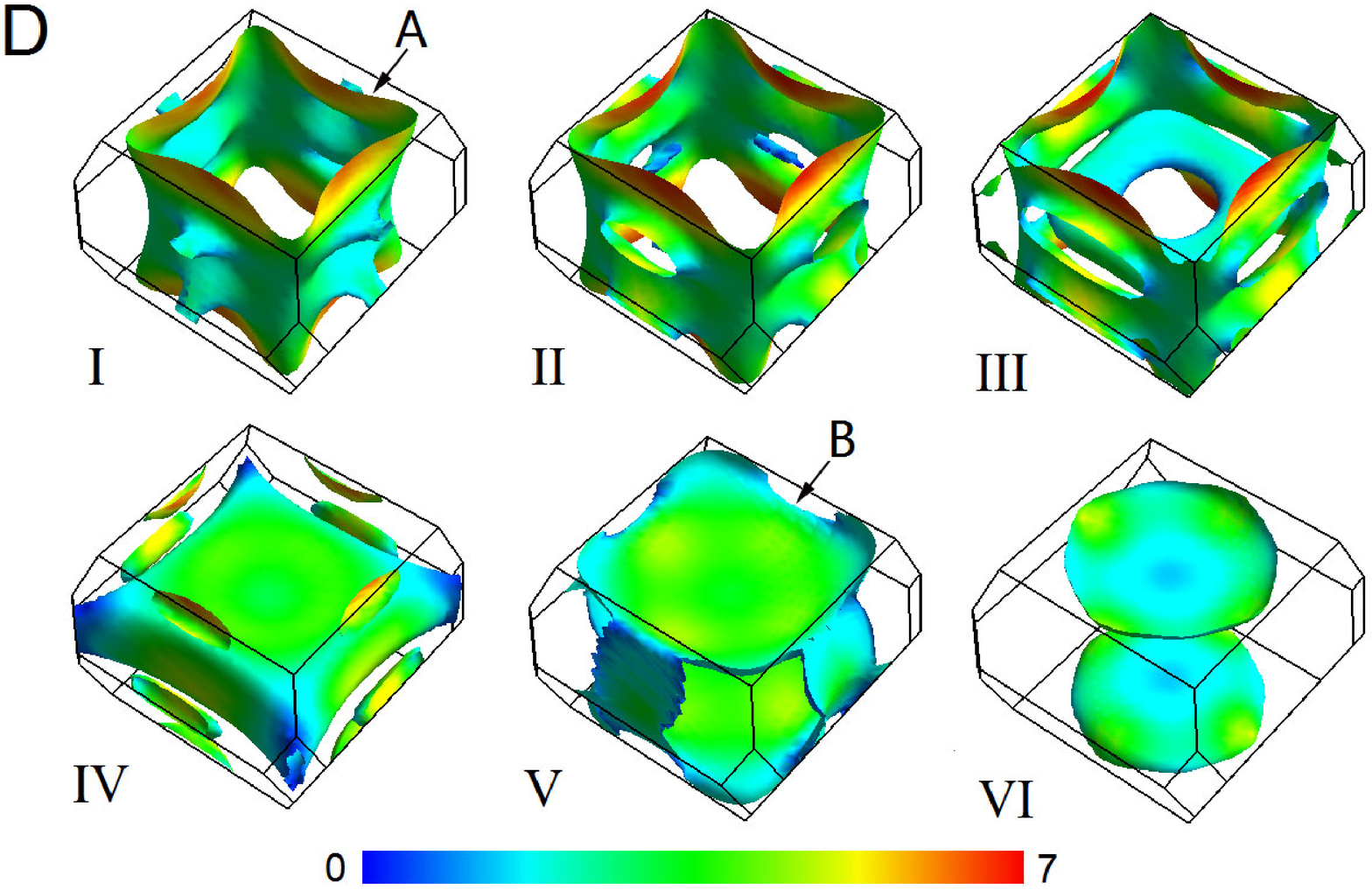}
\caption{(A) Band structure of rocksalt CdO and (B) tetragonal Mn$_2$Au calculated using mBJ potential. (C) Density of states (DOS) of the bulk and interfacial Mn atoms with positive magnetic moment of the ideal Mn$_2$Au$/$CdO(10)$/$Mn$_2$Au multilayer. Inset of (C): orbit dependent DOS of the interfacial Mn atom. In (A), band with index 1, 2, 3 and 4 are flat bands from the $\Gamma$ (0 0 0) to Z (0.5 0.5 -0.5), and 5 from $\Gamma$ to X (0 0 0.5). In (B), band with index 1, 2 and 3 are bands from the $\Gamma$ to X (0.5 0 0.5) point, and 4 from $\Gamma$ to L (0.5 0.5 0.5). (D) Fermi surfaces FSs of tetragonal bulk Mn$_2$Au around Fermi energy $E_F$, $E_{F}-0.3\unit{eV}$, $E_{F}-0.5\unit{eV}$, $E_{F}-1.0\unit{eV}$, $E_{F}-1.5\unit{eV}$, and $E_{F}-1.7\unit{eV}$. The color bar is the Fermi velocity with Rydberg atomic unit. There shows two FSs range from $E_F$ to $E_F-1.7$. The FS A present at $E_F$ and disappearing around $E_F-1.0~\unit{eV}$ is band 5 shown in (A), and the FS B appearing around $E_F-0.3~\unit{eV}$ and disappearing around $E_F-1.7~\unit{eV}$ is band 2 shown in (B).}
\label{fig3}
\end{figure*}

Figure \ref{fig2}C(\Rmnum{1}-\Rmnum{4}) gives the $k_{||}$ resolved spin transmission of the ideal Nb$/$Mn$_2$Au(24)$/$CdO(10)$/$Mn$_2$Au(24)$/$Nb AF-MTJs with S1 structure at the Fermi energy. For the P (AP) magnetic state of the "Nb lead" junctions, the sum of the $k_{||}$ resolved spin transmission of the majority ($\uparrow$) and minority ($\downarrow$) spin is $5.75(0.116) \times 10^{-4}~\unit{e}^{2}/\unit{h}$ and $2.29(0.117) \times 10^{-4}~\unit{e}^{2}/\unit{h}$, respectively. That is, the both spin channels contribute almost equally to the spin transmission of the P magnetic states. This is noticeably different from that in the F-MTJs with considerable MR effect, where the spin transmission is spin polarized.\cite{butler2001} Looking at the 2D Brillouin zone (BZ)(Fig. \ref{fig2}C(\Rmnum{1}) and (\Rmnum{2})), it clearly that the hot spots and hot lines dominate the total transmission for both spin channels, and they are highly spin polarized and locate in different regions. The sum of resonant $k_{||}$ points with transmission possibility larger than $0.01$ contribute to about $94\%$ and $40\%$ of the total spin transmission of the P and AP magnetic state of the "Nb lead" junction. 
About one-fourth of the 2D BZ contributes to the Sharvin conductance of bulk Mn$_2$Au at the Fermi energy, as shown in Fig. \ref{fig2}C(\Rmnum{10}), and we name the region as Sharvin area, which can be understood by the projection of Fermi surface (Fig .\ref{fig3}D(\Rmnum{1})) along the [001] direction and indicate Mn$_2$Au a $k_{||}$ resolved barrier materials. The $k_{||}$ points out of the Sharvin area of Mn$_2$Au contribute to about $67\%$ and $7\%$ of the total transmission for the $\uparrow$ and $\downarrow$ spin of the P magnetic state, respectively. It is less than $1\%$ for both spin channels for the AP state. 
The resonance $k_{||}$ points within and out of the Sharvin area of Mn$_2$Au show difference in details. Both the bonding and antibonding peaks (with energy gap around $10^{-1}\unit{eV}$) of the former are splitting further into two peaks with energy gap around $10^{-4}\unit{eV}$. This maybe related with the complex coupling of the interfacial states.

The $k_{||}$ resolved spin transmission in the "Mn$_2$Au lead" junctions is given in Fig. \ref{fig2}C(\Rmnum{5}-\Rmnum{8}). Similar to the "Nb lead" junctions, the bright areas of the $\uparrow$ and $\downarrow$ spin channels of the P magnetic state of the junction are located in different regions of the 2D BZ also, and the spin transmission is not only spin dependent but also $k_{||}$ dependent. The sum of spin transmission give a larger spin polarization of $9\%$ compared with that in the "Nb lead" junction.

\subsection{Band structure of the system}

N\'u\~nez et al.,\cite{nunez2006theory} gives a theoretical model to estimate the MR effect in the L-type AFM with MR up limit about one hundred percent. Beyond the theoretical model, a more direct way to understand would be the band structure. Figure \ref{fig3} give the band structure of bulk Mn$_2$Au and CdO, the density of states (DOS) and Fermi surface of Mn$_2$Au. At the Mn$_2$Au$/$CdO interface with O bonded with Mn (we can deduce the formation of Mn-O bond from the Mn-O distance around $1.93~\AA$), the former would change the DOS of the latter.
Compared with the DOS of Mn atom in the bulk Mn$_2$Au, the DOS peaks of the $\uparrow$ spin of the $d$ states of the interfacial Mn atom is downward shift, while that of the $\downarrow$ spin upward shift and shrink greatly. The change lead to enhanced spin polarization of the interfacial Mn atoms at the Fermi energy. The orbit dependent DOS of
interfacial Mn atom shown in the inset of Fig. \ref{fig3}C indicates that the $p$ and $d$ orbits contribute mainly to the $\uparrow$ spin, and the total DOS of the $\uparrow$ spin is about four fold larger than that of the $\downarrow$ spin. According to Julliere's model\cite{julliere1975tunneling}, the MR should be around $250\%$, which is close to the calculated value in the AF-MTJs with S1 structure with thinner CdO barrier but considerable smaller than that with thicker barriers.

For the $k_{||}$ resolved barrier nature of tetragonal Mn$_2$Au, the barrier thickness of Nb$/$Mn$_2$Au(24) $/$CdO(10)$/$Mn$_2$Au(24)$/$Nb AF-MTJ is $k_{||}$ dependent also. It is about $2.4$ and $9.1~\unit{nm}$ for the $k_{||}$ points within and out of the Sharvin area of Mn$_{2}$Au, respectively. From the band structure (Fig. \ref{fig3}A), we can find that $k_{||}$ points within or close to the Sharvin area of Mn$_2$Au can satisfy the resonance condition, which maybe responsible for the giant TMRs found in the AF-MTJs with thicker CdO barrier. 

\begin{figure}[tbp]
\centering
\includegraphics[width=8.6cm]{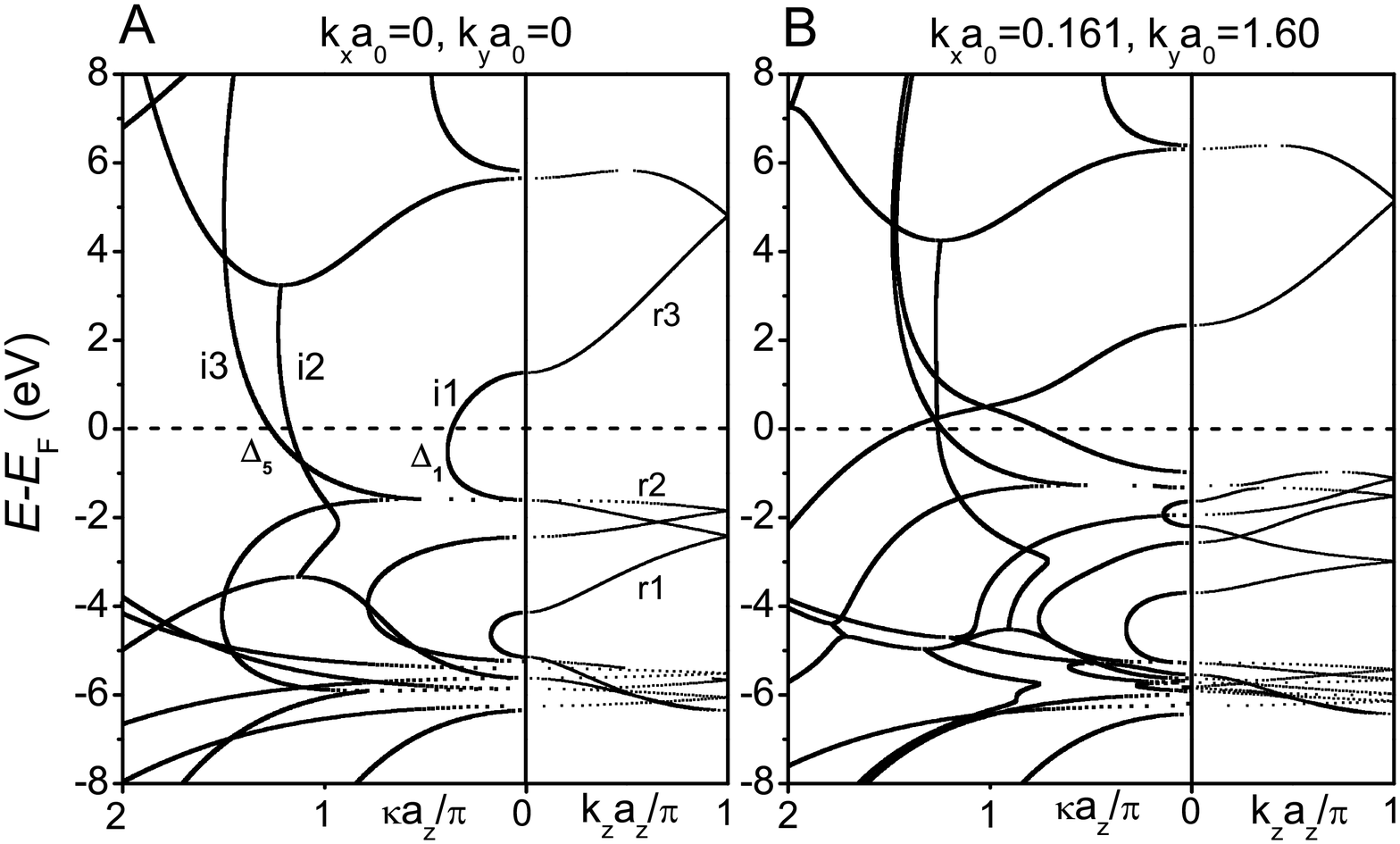}
\includegraphics[width=8.6cm]{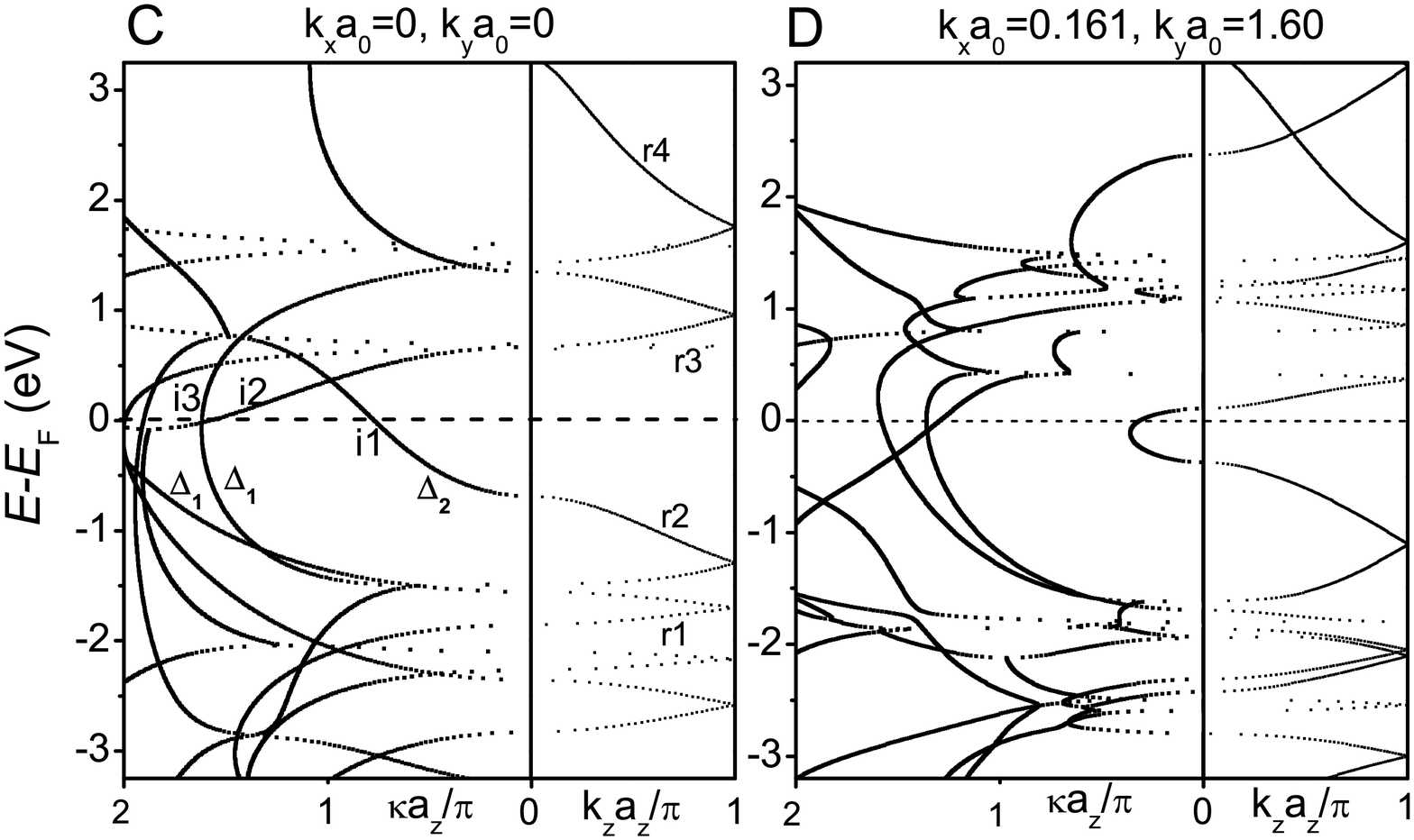}
\caption{Complex band structure of the sandwiched (a-b) CdO and
(c-d) Mn$_2$Au in the Mn$_2$Au$/$CdO$/$Mn$_2$Au AF-MTJs at 2D $\Gamma$ and one resonance $k_{||}$ points. Therein, $k_z$ and $\kappa$ are real and imaginary part of wave vector $k$,
$a_0=3.328~\unit{nm}$ is the lateral lattice parameter of the
scattering region, and $a_z$ is the thickness of the repeat unit along the
transport direction, which is 4.69 and 8.539 $\unit{nm}$ for CdO and
Mn$_2$Au, respectively. The real part bands are folding. Bands with index r1, r2, and r3 in (a) are corresponding to the bands with index 1, 2, and 3 in Fig. \ref{fig4}(c), respectively, and Bands with index r1, r2, r3 and r4 in (c) are corresponding to the bands with index 1, 2, 3, and 4 in Fig. \ref{fig4}(b), respectively.} \label{fig4}
\end{figure}

Complex band structure can be used to estimate the tunnel transmission through the barrier. Figure \ref{fig4} gives the complex band structure of bulk CdO and Mn$_2$Au at 2D
$\Gamma$ point and a resonant $k_{||}$ point along the [001] direction. Here, we pay attention to the complex band with smaller imaginary part in the vicinity of Fermi energy. The symmetry of the imaginary bands can be identified by the symmetry of the connected real bands. Rocksalt CdO has crystal structure as MgO, and the complex band structure of CdO is similar to that of MgO. In Fig. \ref{fig4}A, we can identify a $\Delta_1$ imaginary band connecting with a $s$ band (band r3 of Fig. \ref{fig4}A) and a $d_{z{^2}}$ band (band r1 of Fig. \ref{fig4}A) at $\Gamma$ point, and a $\Delta_5$ imaginary band connecting with a doubly degenerate bands $d_{zx}$ and $d_{zy}$ (band r2 of Fig. \ref{fig4}A). Between the $\Delta_1$ and $\Delta_5$ imaginary bands, the symmetry of the imaginary band i2 is hard to identify. The $\Delta_1$ imaginary band shows smaller $\kappa$ ($k=k_z+i\kappa$) in the vicinity of the Fermi energy, which would dominate the tunneling transmission, indicating the potential symmetry filtering application as found in the Fe$/$MgO$/$Fe junction\cite{butler2001}. Furthermore, semiconductor type band gap and small effective mass at $\Gamma$ point (Fig. \ref{fig3}B) is another favorable character for the spintronic application.

The symmetry of the imaginary bands of tetragonal Mn$_2$Au is complex. There presents two parabolic imaginary bands with evident $\Delta_1$ symmetry and several irregular imaginary bands at the 2D $\Gamma$ point. The $\Delta_1$ imaginary band i3 is connected with band r2 and r4 at $Z$ point, and another $\Delta_1$ imaginary band (with larger $\kappa$) connected with band r1 and r3 at $\Gamma$ point, as shown in Fig. \ref{fig4}C. To identify the symmetry of the irregular band, especially the one (band i1) with smallest $\kappa$, we studied the Fermi surface (FS) around and beyond the Fermi energy as shown in Fig \ref{fig3}D. Therein, the symmetry of both FS A and B are energy dependent. At energy $E_F-0.3~\unit{eV}$, the FS B shows four fragments sited around $\phi=$ 0, 90, 180 and 270 degrees with $d_{xy}$ (or $d_{x^2-y^2}$) symmetry, which form an onion rings structure around $E_{F}-0.5\unit{eV}$, and then turn into thin square drum with small gaps at four corners around $E_{F}-1.0\unit{eV}$, and then the square drum gets thicker and the gaps expands until two sheets sited parallel to the $xy$ face formed around $E_{F}-1.7\unit{eV}$. Within energy range from $E_{F}-1.0\unit{eV}$ to $E_{F}-1.7\unit{eV}$, The FS B can be identified as a combination of $p_z$ and $d_{z^2}$ symmetry. For the imaginary band i1 is connected with a $d_{xy}$ orbit, which can be identified as a $\Delta_2$ symmetry. 

At Fermi energy, the electron is predicted to decay with the rate exp$(-2\kappa \Delta_z)$, where 
$\Delta_z$ is the thickness of barrier. According to the symmetry filtering scheme,\cite{butler2001} bands can and only can accommodate incoming states with same symmetry as itself. Firstly, we pay attention to the $\Delta_1$ symmetry states. Along the transport direction, we can estimate that $\kappa a_z$ is about $0.37\pi$ for the $\Delta_1$ electrons of CdO, and $1.62\pi$ for the $\Delta_1$ electrons of Mn$_2$Au. Supposing a symmetric AF-MTJ with 24Ls Mn$_2$Au and 10Ls CdO, the tunneling transmission of the $\Delta_1$ states at the 2D $\Gamma$ point would be about $4.1\times10^{-41}~\unit{e}^{2}/\unit{h}$. Supposing the imaginary band i2 of CdO follows same symmetry as the imaginary band i1 of Mn$_2$Au, the former has $\kappa a_z$ of $1.11\pi$ and the latter $0.67\pi$. The spin transmission in the AF-MTJs carried by the $\Delta_2$ state would be around $1.7\times10^{-30}~\unit{e}^{2}/\unit{h}$. Comparatively, the transmission in the ideal Nb$/$Mn$_2$Au(24)$/$CdO(10)$/$Mn$_2$Au(24)$/$Nb AF-MTJs with S1
structure calculated based on the scattering wave function is about
$1.0\times10^{-49}$ and $3.7\times10^{-51}~\unit{e}^{2}/\unit{h}$
for the P and AP magnetic states, respectively. Our calculations indicate that the $\Delta_1$ imaginary band should be responsible for the spin transmission the 2D $\Gamma$ points. Many effects should be responsible for the giant difference between the model and the first-principles calculations. The leading factors should be band mismatch, the interface scattering, the electron mass entering into the barriers, and so on.

For $k_{||}$ points slightly away the 2D $\Gamma$ points, the average spin transmission calculated first-principally of the $k_{||}$ points within $|k_x|\le\pi/100$ and $|k_y|\le\pi/100$ is about
$1.2\times10^{-29}$ and $2.3\times10^{-33}~\unit{e}^{2}/\unit{h}$
for the P and AP magnetic states of the ideal Nb$/$Mn$_2$Au(24)$/$CdO(10)
$/$Mn$_2$Au(24)$/$Nb AF-MTJs with S1 structure, respectively. The calculated
numbers are about twenty orders in magnitude larger than that calculated at the 2D $\Gamma$ point, which is about ten orders in magnitude larger than the model estimation carried by the $\Delta_1$ states, but close to the model estimation carried by the $\Delta_2$ states (supposing the imaginary band i2 of CdO follows a $\Delta_2$ symmetry). 

As the $k_{||}$ points walk outward from the 2D $\Gamma$ point, the $k_{||}$ dependent barrier height and band gap of rocksalt CdO would decrease for the join of the band from $\Gamma$ to X (band 4 in Fig. \ref{fig3}B). Similar effect is found in tetragonal Mn$_2$Au, also. When the interfaces in the AF-MTJs are clean enough, some $k_{||}$ points with appropriate barrier height maybe meet the resonant tunneling condition. For the ideal "Nb lead" AF-MTJs, the barrier is so thick that only the $k_{||}$ points within or close to the Sharvin area of Mn$_2$Au can meet the resonant tunneling conditions. For a resonant $k_{||}$ points within the Sharvin area of Mn$_2$Au with
$k_x a_0=0.161$ and $k_y a_0=1.60$ as shown in Fig. \ref{fig4}B
and D, the imaginary part of the wave vector contribute by the $d$ states are considerable
smaller than that at the 2D $\Gamma$ point as shown in Fig. \ref{fig4}A and C, respectively. That is, the interfacial resonant tunneling is directly responsible for the giant TMRs effect in the pure AF-MTJs with thicker CdO, but the $k_{||}$ dependent band structure of the system is more fundamental.

\begin{figure}[tbp]
\centering
\includegraphics[width=8.6cm]{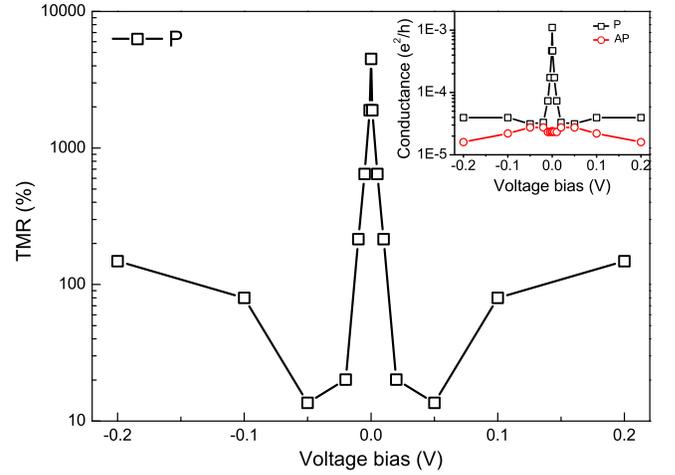}
\caption{Voltage bias dependent TMRs and conductances in the ideal
Nb$/$Mn$_2$Au(24)$/$CdO(10)$/$Mn$_2$Au(24)$/$Nb pure AF-MTJs with S1
structure.} \label{fig5}
\end{figure}

\subsection{Voltage bias dependence of TMRs}
Another effect of the complex band structures in the "Nb lead" AF-MTJs would be the voltage bias dependent TMR effect as shown in Fig. \ref{fig5}. When voltage bias is finite, the potentials in the CdO region shift linearly while that in the Mn$_2$Au region keep unchange for the metallicity, which would affect the interfacial resonance.\cite{ke10} So, the spin transmissions in the ideal Nb$/$Mn$_2$Au(24) $/$CdO(10)$/$Mn$_2$Au(24)$/$Nb AF-MTJs would be sensitive to the voltage
bias. A voltage bias of $0.001\unit{eV}$ can reduce the spin
transmission in the P magnetic state of the AF-MTJs with S1 structure by a factor of twenty as shown in the inset of Fig. \ref{fig5}. Comparatively, the total spin transmission of the AP magnetic state of the AF-MTJs is not so sensitive to the
voltage bias. As a result, the TMR of the AF-MTJs junction
decrease firstly from $4500\%$ at equilibrium state to $14\%$ at
voltage bias of $0.05~\unit{eV}$, and then turn to increase to $150\%$ at
$0.2~\unit{eV}$.

\begin{table*}[tbp]
\caption{Tunneling conductances and TMRs of the ideal
Nb$/$Mn$_2$Au(24)$/$CdO(10)$/$Mn$_2$Au(24)$/$Nb(001) pure AF-MTJs at zero
and $0.1~\unit{eV}$ (in the bracket) voltage bias. The unit of
conductance is $10^{-5}~\unit{e}^{2}/\unit{h}$.}
\label{tab1}%
\begin{tabular*}{17.6cm}{@{\extracolsep{\fill}}lccccc}
\hline\hline
Structures & P &  AP &PP& TMR(P) $(\%)$ & TMR(PP) $(\%)$  \\
\hline
S1  & 110(3.96)  & 2.35(2.20)&77.2(3.16) & 4580(80)& 3180(44) \\
S2 & 4.28(1.36)  & 0.180(0.101) &2.27(0.737) &2270(1250) &1160(536) \\
S3 &   0.925(0.0667)  &   0.011(0.0191)&0.49(0.0429) &8300(250) & 4350(125) \\
A1 &  0.468(0.0369) &   0.279(0.0230)  & 0.375(0.0301)&68(60) & 34(31)\\
A2 & 0.0583(0.0250) & 0.0348(0.0136) &0.0466(0.0199) &67(84) &34(46)\\
A3 &  0.967(1.55)& 0.68(0.755)  &0.773(1.19) &42(105) & 14(57)
\\\hline
S1$^1$  &  9.15 &2.59&-& 252&- \\
S1$^2$  & 28.9 &12.9 &-& 123&-  \\
S1$^3$  & 10.4& 2.61  &-&298& -\\
\hline\hline
\multicolumn{6}{l} {$^1$ $10\%$ Oxygen vacances at both Mn$_2$Au$/$CdO interfaces}    \\
\multicolumn{6}{l} {$^2$ $10\%$ Manganese vacances at both Mn$_2$Au$/$CdO interfaces}    \\
\multicolumn{6}{l} {$^3$ $10\%$ Manganese-Cadmium (Mn-Cd) exchanges at both Mn$_2$Au$/$CdO interfaces}    \\
\end{tabular*}%
\end{table*}

Table \ref{tab1} summarizes the conductances and TMRs of the
ideal Nb$/$Mn$_2$Au(24)$/$CdO(10)$/$Mn$_2$Au(24)$/$Nb(001) AF-MTJs
at zero and $0.1~\unit{eV}$ voltage bias. At equilibrium state, the
symmetric structures show larger tunneling conductances and TMRs
than the asymmetric ones, and the S1 structure shows largest
tunneling conductance among all studied cases, and the S3 structure
shows largest TMRs among all studied cases. In the presence of
voltage bias of $0.1~\unit{eV}$, the tunneling conductances of the P
and PP magnetic states of the symmetric structures S1 and S3 are more
sensitive to the voltage bias, while the AP structure of which are
insensitive to the voltage bias, leading to sharp decrease of the
TMR effect. The S2 structure is an exception among the symmetric
structures, both the tunnel conductances and TMRs are insensitive to
the finite voltage bias. The P, PP and AP magnetic states of
the asymmetric structures A1 and A2 show similar effect in the
presence of voltage bias, the TMRs in the two structure is
insensitive to the voltage bias also. The A3 structure is an
exception among the asymmetric structures, both the tunneling
conductances and TMRs increase under finite voltage bias.
Summarily, the tunneling conductance and TMR effect in the ideal "Nb lead" AF-MTJs show complex dependence on the magnetic structures, the atomic structures, and the voltage bias. These complex dependence can be understood by the complex band structure of
the CdO and Mn$_2$Au in companion with the localized magnetism at the
Mn$_2$Au$/$CdO interface induced by interfacial effect as
discussed above.

\subsection{Interfacial disorder}
Ideal heterostructure is hard to realize experimentally, many kinds imperfects present and concentrate at the interfaces. Here, we pay attention to three kinds imperfects as shown in the Table \ref{tab1}. Generally, the introductions of interfacial disorder would deteriorate the interfacial resonance tunneling effect between the same spin channels while enhance the scattering probability among the different spin channels.\cite{ke10} Compared with interfacial Oxygen vacancy and Manganese-Cadmium (Mn-Cd) exchanges with same concentration, the interfacial Manganese vacancy enhance the scattering probability among the different spin channels more. The presence of $10\%$ interfacial imperfects would reduce the TMRs in the Nb$/$Mn$_2$Au(24)$/$CdO(10)$/$Mn$_2$Au(24)$/$Nb(001) AF-MTJs to around one or several hundreds percent at equilibrium state. To achieve larger TMR effect in the AF-MTJs, the interfaces should be as clean as possible.

\section{Summary}
Summarily, we calculate the spin transmission and TMR effect in the Mn$_2$Au-based pure
AF-MTJs based on a first-principle scattering theory. Giant TMRs with
order of $1000\%$ are found in some symmetric junctions while that around $100\%$
found in some asymmetric junctions. The interfacial resonance
tunneling effect related with the $k_{||}$ dependent band structures
of the system and enhanced magnetism of the interfaces
atoms should be responsible for the giant TMR effect therein. The
TMR effect in the ideal AF-MTJs is sensitive to the interfacial structure,
voltage bias, and interfacial disorder. For a
ideal symmetric junction with S1 structure with 24 Ls Mn$_2$Au and 10 Ls CdO, a
voltage bias of $0.1~\unit{eV}$ would reduce the TMR from $4580\%$
at equilibrium state to $80\%$, and the introduction of $10\%$
interfacial disorder such as O vacancy, Mn vacancy, and Mn-Cd
exchanges at the Mn$_2$Au$/$CdO interfaces would reduce the TMR
 to the order of one to several hundreds percent at the equilibrium state. The giant TMR effect predicted in the pure AF-MTJs indicates the possibility to exploit the  antiferromagnetism without the aid of ferromagnet or ferrimagnet. Moreover, good symmetry filtering effect in companion with the semiconductor-type band gap suggests rocksalt CdO a promising material for spintronic applications.

\section*{Acknowledgement}

We gratefully acknowledge financial support from the National
Natural Science Foundation of China (Grant No. 12074102, 11804062 and
11804310).

\bibliographystyle{apsrev}
\bibliography{db}

\end{document}